\documentstyle[12pt]{article}
\newlength{\extraspace}
\newlength{\extraspaces}
\newcommand{\be}{\begin{equation}
\addtolength{\abovedisplayskip}{\extraspaces}
\addtolength{\belowdisplayskip}{\extraspaces}
\addtolength{\abovedisplayshortskip}{\extraspace}
\addtolength{\belowdisplayshortskip}{\extraspace}}
\newcommand{\ee}{\end{equation}}
\newcommand{\ba}{\begin{eqnarray}
\addtolength{\abovedisplayskip}{\extraspaces}
\addtolength{\belowdisplayskip}{\extraspaces}
\addtolength{\abovedisplayshortskip}{\extraspace}
\addtolength{\belowdisplayshortskip}{\extraspace}}
\newcommand{\ea}{\end{eqnarray}}
\newcommand{\newsection}[1]{
\vspace{7mm}
\pagebreak[3]
\addtocounter{section}{1}
\setcounter{equation}{0}
\setcounter{subsection}{0}
\setcounter{footnote}{0}
\begin{center}
{\large {\bf \thesection. #1}}
\end{center}
\nopagebreak
\medskip
\nopagebreak
\hspace{3mm}}
\newcommand{\nonu}{\nonumber \\[.5mm]}
\newcommand{\A}{&\!\!\!}

\begin{document}
\addtolength{\baselineskip}{.7mm}
\begin{flushright}
STUPP--95--138 \\
January, 1995
\end{flushright}
\vspace{2mm}
\begin{center}
{\large{\bf{Ashtekar Variables and Matter Coupling}}}
\footnote{\ Talk at the workshop on ``General Relativity and Gravitation''
held at Kyoto University during Nov.\ 28-Dec.\ 1, 1994.} \\[3mm]
{\sc Motomu Tsuda, Takeshi Shirafuji and Hong-jun Xie} \\[1mm]
{\it Physics Department, Saitama University \\[1mm]
Urawa, Saitama 338, Japan} \\[8mm]
{\bf Abstract}\\[2mm]
{\parbox{13cm}{\hspace{5mm}
It has been shown for low-spin fields that the use of only the self-dual
part of the connection as basic variable does not lead to
spurious equations or inconsistencies. We slightly generalize
the form of the chiral Lagrangian of half-integer spin fields
and express its imaginary part in a simple form. If the imaginary
part is non-vanishing, it will lead to spurious equations.
As an example, for (Majorana) Rarita-Schwinger fields the equations
of motion of the torsion is solved and it is shown that it vanishes
owing to the Fierz identity.}}
\end{center}
\setcounter{section}{0}
\setcounter{equation}{0}
\newsection{Introduction}

In the mid-1980s Ashtekar has presented a new formulation
of general relativity from a non-perturbutive point of view,
in terms of which all the constraints of the gravity become
simple polynomials of the canonical variables \cite{AA}-\cite{AAL}.
This formulation can be extended to include matter sources.
In particular, in the cases of spin-1/2 fields \cite{AAL,ART}
and $N = 1$ supergravity \cite{AAN,JJ}, the constraints are again
polynomials of the canonical variables. On the other hand,
the Ashtekar formulation of $N = 1$ supergravity was reformulated
in the method of the two-form gravity \cite{CDJ}.
The progress of the Ashtekar formalism can be traced from
the references compiled in \cite{Bib.}.

Here the complex chiral action using the self-dual connection
and the tetrads $e^i_\mu$ is
\be
S^{(+)} = \int d^4 x \, e R^{(+)} + [{\rm matter \ terms}], \label{action+1}
\ee
where $e = {\rm det}(e^i_\mu)$ and
\be
R^{(+)} := {1 \over 2} \left(R - {i \over 2} {\epsilon^{ij}}_{\! kl}
                  {R^{kl}}_{\! \mu \nu} e^\mu_i e^\nu_j \right).
\ee
Greek letters $\mu, \nu, \cdots$ are space-time indices,
and Latin letters {\it i, j,} $\cdots$
are local Lorentz indices. Using the Bianchi identity,
the chiral action (\ref{action+1}) is reexpressed as
\be
S^{(+)} = {1 \over 2} \int d^4 x \, e R(e)
+ [{\rm quadratic \ terms \ of \ torsion}]
+ [{\rm matter \ terms}]. \label{action+2}
\ee
The first term of (\ref{action+2}) is the Einstein-Hilbert action.
Since the torsion equals zero in the source-free case,
the complex chiral action is equivalent to the Einstein-Hilbert action.
In this paper we shall take the form of matter terms to be slightly
general, and analyze the consistency of the field equations.

In Sec.2 we will contemplate the half-integer spin fields
minimally coupled to gravity, and introduce a slightly general
form of the complex chiral Lagrangian of matter fields.
In Sec.3 we will argue if the field equations are influenced by
the imaginary part of the chiral Lagrangian.
In the final section our result is summarized.

\newsection{Lagrangian of Matter Fields}

In the source-free case, although the chiral gravitational
Lagrangian is complex, the imaginary part of the Lagrangian turns out
to be simply the Bianchi identity.
Since the Lagrangian of integer spin fields (spin-0,1) does not
contain the Lorentz connection, the situation remains unaltered.
Therefore, let us consider half-integer spin fields
($\psi$ and $\psi_\mu$ for spin-1/2 and spin-3/2 fields, respectively)
as matter fields.

Firstly, let us suppose that the matter Lagrangian is obtained
by the minimal prescription; namely, by replacing ordinary
derivatives by covariant derivatives,
\be
\partial_i \rightarrow e_i^\mu D_\mu
\ee
with
\be
D_\mu = \partial_\mu + {i \over 2} A_{ij \mu} S^{ij}.
\ee
Here $e_i^\mu$ is a tetrad field, $A_{ij \mu}$ means
the Lorentz connection, and $S_{ij}$ stands for the SL(2,{\bf C}) generator.
\footnote{\ In our convention $S_{ij} = {i \over 4}[\gamma_i, \gamma_j]$
and $\{ \gamma_i, \gamma_j \} = - 2 \eta_{ij}$. We denote the Minkowski
metric by $\eta_{ij}$ = diag(-1, +1, +1, +1).
$\epsilon_{\mu \nu \rho \sigma}$ is the totally antisymmetric tensor
normalized as $\epsilon_{0123} = + e$.}
The Lorentz connection $A_{ij \mu}$ is divided into
the Ricci rotation coefficients $A_{ij \mu}(e)$
and contorsion tensor $K_{ij \mu}$,
\be
A_{ij \mu} = A_{ij \mu}(e) + K_{ij \mu}.
\ee

Secondly, we suppose that the chiral Lagrangian of matter fields
is described by using the self-dual part of the Lorentz connection.
According to the equation
\be
A^{(+)}_{ij \mu} S^{ij} = A_{ij \mu} S^{ij} {1 + \gamma_5 \over 2},
\ee
this demand that only the terms expressed by $D_\mu \psi_R$
and $\overline \psi_L \overleftarrow D_\mu$ should appear
in the matter Lagrangian. $\psi_R$ ($\psi_L$) is
the right (left)-handed spinor field;
\ba
\left\{ \matrix{\psi_R \A = \A \displaystyle {{1 + \gamma_5 \over 2}}
    \psi, \vspace{1mm} \cr
       \psi_L \A = \A \displaystyle {{1 - \gamma_5 \over 2}}
    \psi. \cr } \right.
\ea

Let us consider a (Majorana) Rarita-Schwinger field. (Spin-1/2 fields
can be considered in the similar manner.)
Its Lagrangian in Minkowski space is
\be
L_{RS} = {1 \over 2} \epsilon^{\mu \nu \rho \sigma}
              \overline \psi_\mu \gamma_5 \gamma_\rho
              \partial_\sigma \psi_\nu. \label{LRS-1}
\ee
Applying the minimal prescription to (\ref{LRS-1}),
the Lagrangian density of a Rarita-Schwinger field becomes
\be
{\cal L}_{RS} = {1 \over 2} e \epsilon^{\mu \nu \rho \sigma}
               \overline \psi_\mu \gamma_5 \gamma_\rho
               D_\sigma \psi_\nu, \label{LRSE}
\ee
which is used in $N = 1$ supergravity \cite{FNF}-\cite{NWH}.

On the other hand, the Lagrangian (\ref{LRS-1}) can be rewritten as
\be
L_{RS} = - \epsilon^{\mu \nu \rho \sigma} \overline \psi_{R \mu}
              \gamma_\rho \partial_\sigma \psi_{R \nu}
              + [{\rm tot.\ div.\ term}]. \label{LRS-2}
\ee
The minimal prescription applied to (\ref{LRS-2}) leads to
the chiral Lagrangian density,
\be
{\cal L}^{(+)}_{RS} = - e \epsilon^{\mu \nu \rho \sigma}
                     \overline \psi_{R \mu} \gamma_\rho
                     D_\sigma \psi_{R \nu}. \label{LRSE+}
\ee
Here the total divergence term in (\ref{LRS-2}) is discarded.
(The chiral Lagrangian density for the antiself-dual connection
can be similarly defined by using the left-handed spinor field.)

The Lagrangian densities (\ref{LRSE}) and (\ref{LRSE+}) can respectively
be reexpressed as
\ba
{\cal L}_{RS} \A = \A {\cal L}_{RS}(e) + {i \over 4}e
                  \epsilon^{\mu \nu \rho \sigma} \overline \psi_\mu
                  \gamma_5 \gamma_\rho K_{ij \sigma} S^{ij} \psi_\nu,
                  \label{LRSK} \\
{\cal L}^{(+)}_{RS} \A = \A {\cal L}_{RS}(e) + {i \over 2}e
                  \epsilon^{\mu \nu \rho \sigma} \overline \psi_\mu
                  \gamma_5 \gamma_\rho K^{(+)}_{ij \sigma} S^{ij} \psi_\nu,
                  \label{LRSK+}
\ea
where $K^{(+)}_{ij \mu}$ is the self-dual part of contorsion tensor
$K_{ij \mu}$. In (\ref{LRSK+}) a total divergence term has been omitted.
We notice that the factor of the second term in (\ref{LRSK+}) is
twice that in (\ref{LRSK}).

Let us consider matter fields of which the real Lagrangian density
is written as
\be
{\cal L}_M = {\cal L}_M(e) + eX_{ijk} K^{ijk}, \label{LMK}
\ee
where $X_{ijk} = X_{[ij] k}$ is a real tensor made of matter fields.
Based on the case for spin-3/2 (and spin-1/2) fields,
we assume that the complex chiral Lagrangian density
of those matter fields is expressed as follows;
\be
{\cal L}^{(+)}_M = {\cal L}_M(e) + 2eX_{ijk} K^{(+) ijk}. \label{LMK+}
\ee
It is noted that only the self-dual part $X^{(+)}_{ijk}$ contributes
in (\ref{LMK+}).

In order to analyze the Lagrangian densities (\ref{LMK}) and (\ref{LMK+}),
it is convenient to decompose $K_{ijk}$ into irreducible parts \cite{HS};
\footnote{\ The $u_{ijk}$ of (\ref{K}) is related to $t_{ijk}$
of Ref. \cite{HS} by $u_{k (ij)} = t_{ijk}$.}
\be
K_{ijk} = u_{ijk} - {2 \over 3} \eta_{k [i} v_{j]}
             + {1 \over 2} \epsilon_{ijkl} a^l, \label{K}
\ee
where the tensor $u_{ijk} = u_{[ij] k}$ is traceless and
\be
\epsilon^{ijkl} u_{jkl} = 0.
\ee
Here the vector $v_i$ is
\be
v_i = {K_{ij}}^j,
\ee
and the axial vector $a_i$ is
\be
a_i = {1 \over 3} \epsilon_{ijkl} K^{jkl}.
\ee
In the same way, $K^{(+)}_{ijk}$ is decomposed as
\be
K^{(+)}_{ijk} = u^{(+)}_{ijk} - {2 \over 3} \eta_{k [i} v^{(+)}_{j]}
                  + {1 \over 2} \epsilon_{ijkl} a^{(+) l}, \label{K+}
\ee
where $u^{(+)}_{ijk}$ is self-dual part of $u_{ijk}$, and
\be
a^{(+)}_i = {2 \over 3}i v^{(+)}_i = {1 \over 2}
\left(a_i + {2 \over 3}i v_i \right).
\ee

Substituting (\ref{K}) and (\ref{K+}) into (\ref{LMK}) and (\ref{LMK+}),
respectively, the matter Lagrangian densities can be represented as
\ba
{\cal L}_M \A = \A {\cal L}_M(e) + e(B_{ijk} u^{ijk}
                              + C_i v^i + D_i a^i), \label{LMu} \\
{\cal L}^{(+)}_M \A = \A {\cal L}_M(e) + 2e(B_{ijk} u^{(+) ijk}
                              + C_i v^{(+) i} + D_i a^{(+) i}), \label{LMu+}
\ea
where $B_{ijk}=B_{[ij] k}, C_i$ and $D_i$ are related to $X_{ijk}$.
We notice that only the self-dual part $B^{(+)}_{ijk}$ contributes
in (\ref{LMu+}).

\newsection{Consistency of the field equations}

Let us write the quadratic terms of the torsion
in (\ref{action+2}) explicitly. The gravitational Lagrangian density
${\cal L}_G = eR$ is represented as follows;
\be
{\cal L}_G = {\cal L}_G(e) + {e \over 2} \left({1 \over 2}u^2
             - {2 \over 3}v^2 + {3 \over 2}a^2 \right), \label{LGu}
\ee
where $u^2 = u^{ijk} u_{ijk}$, etc. We define the total Lagrangian
density as the sum of (\ref{LGu}) and (\ref{LMu});
\ba
{\cal L} = {\cal L}(e)
       + \displaystyle {{e \over 2} \left({1 \over 2}u^2
                        - {2 \over 3}v^2 + {3 \over 2}a^2 \right)}
       + e(B_{ijk} u^{ijk} + C_i v^i + D_i a^i).
\label{totL}
\ea
On the other hand, the chiral gravitational Lagrangian
${\cal L}^{(+)}_G = e R^{(+)}$ of (\ref{action+1}) becomes
\be
{\cal L}^{(+)}_G = {\cal L}_G(e) + e \left({1 \over 2}u^{(+)2}
        - {2 \over 3}v^{(+)2} + {3 \over 2}a^{(+)2} \right), \label{LGu+}
\ee
and the total chiral Lagrangian density is
\ba
{\cal L}^{(+)} \A = \A {\cal L}^{(+)}(e)
               + \displaystyle {e \left({1 \over 2}u^{(+)2}
                - {2 \over 3}v^{(+)2} + {3 \over 2}a^{(+)2} \right)} \nonu
         \A \A + 2e(B_{ijk} u^{(+) ijk} + C_i v^{(+) i} + D_i a^{(+) i}),
\label{totL+}
\ea
which is the sum of (\ref{LGu+}) and (\ref{LMu+}).

To begin with, let us consider the equations of motion for
$u_{ijk}, v_i$ and $a_i$. The equations of motion derived from
(\ref{totL}) are
\ba
\left\{ \matrix{ u_{ijk} \A = \A - 2 B_{ijk}, \vspace{1mm} \cr
               v_i \A = \A \displaystyle {{3 \over 2}} C_i, \vspace{1mm} \cr
               a_i \A = \A - \displaystyle {{2 \over 3}} D_i. \cr } \right.
               \label{uva}
\ea
On the other hand, regarding $u^{(+)}_{ijk}$ and $v^{(+)}_i$
as independent variables in (\ref{totL+}), we obtain the equations
of motion as
\ba
\left\{ \matrix{ u^{(+)}_{ijk} \A = \A - 2 B^{(+)}_{ijk},\hfill
\vspace{1mm} \cr
               v^{(+)}_i \A = \A \displaystyle {{3 \over 4}}
                          \left(C_i + \displaystyle {{2 \over 3}}i
                         D_i \right),\hfill \cr } \right. \label{uva+}
\ea
where $B^{(+)}_{ijk}$ is self-dual part of $B_{ijk}$.
It can be shown that $u_{ijk}, v_i$ and $a_i$ derived from (\ref{uva+})
coincide with (\ref{uva}).
If $u_{ijk}, v_i$ and $a_i$ are regarded as independent
variables in ${\cal L}^{(+)}$ of (\ref{totL+}), both Re${\cal L}^{(+)}$
and Im${\cal L}^{(+)}$ give the same result as (\ref{uva});
namely, although ${\cal L}^{(+)}$ is complex, extra conditions do not appear.

Next, let us turn to investigate the detailed form of Im${\cal L}^{(+)}$.
Using the solution (\ref{uva}) in (\ref{totL+}),
Im${\cal L}^{(+)}$ can be written
in terms of $u_{ijk}, v_i$ and $a_i$ as
\be
{\rm Im}{\cal L}^{(+)} = {e \over 8}
                    (\epsilon_{ijmn} u^{ijk} {u^{mn}} \! _k
                    - 8 v_i a^i). \label{Imuva}
\ee
After a little calculation, we get
\be
{\rm Im}{\cal L}^{(+)} = {e \over 8} \epsilon_{ijmn}
                        T^{kij} {T_k}{\! ^{mn}}, \label{ImT}
\ee
where $T_{ijk} = K_{ijk} - K_{ikj}$ is the torsion tensor.

In the case of spin-1/2 fields,
\ba
\left\{ \matrix{ u_{ijk} \A = \A 0,\hfill \vspace{1mm} \cr
                 v_i \A = \A 0,\hfill \vspace{1mm} \cr
                 a_i \A = \A \overline \psi
                 \gamma_5 \gamma_i \psi.\hfill \cr } \right. \label{uva1/2}
\ea
Substituting (\ref{uva1/2}) into (\ref{Imuva}),
we obtain Im${\cal L}^{(+)} = 0$.
For a (Majorana) Rarita-Schwinger field, we have
\ba
\left\{ \matrix{ u_{ijk} \A = \A - \displaystyle {{i \over 4}}
                      {\epsilon_{ij}}^{\! rs} \epsilon_{mnkr}
                      \overline \psi^m \gamma_s \psi^n\hfill
   \vspace{1mm} \cr
                      \A \A - [{\rm trace \ part}]
                      - [{\rm tot.\ antisym.\ part}],\hfill
   \vspace{2mm} \cr
           v_i \A = \A \displaystyle {{i \over 2}} \overline \psi_i
                       \gamma^j \psi_j,\hfill
   \vspace{1mm} \cr
           a_i \A = \A - \displaystyle {{i \over 12}} \epsilon_{ijkl}
                               \overline \psi^j
                       \gamma^l \psi^k.\hfill \cr } \right. \label{uvaRS}
\ea
Although $u_{ijk}$ are considerably complicated, it can be shown
after a little calculation that the torsion takes a simple form
\be
T_{kij} = - {i \over 2} \overline \psi_i \gamma_k \psi_j, \label{torsion}
\ee
which is the same result as $N = 1$ supergravity.
With the help of the following identity
\be
\epsilon^{ijkl} (\overline \psi_k \gamma_m \psi_l)
\gamma^m \psi_j = 0, \label{Fierz}
\ee
which is valid because of the Fierz identity \cite{NWH2},
substituting (\ref{torsion}) into (\ref{ImT}) shows that
Im${\cal L}^{(+)}$ vanishes.

\newsection{Summary}

Based on the considerations of spin-1/2 fields and (Majorana)
Rarita-Schwinger fields, we have generalized the chiral Lagrangian
of half-integer spin fields a little. This generalized
chiral Lagrangian of matter fields holds also in the case of
$N = 2$ supergravity and (Dirac) Rarita-Schwinger fields, etc.
For $N$-(Majorana) Rarita-Schwinger fields, we may merely replace
$\psi_\mu$ in this paper by $\psi_\mu^I$ where $I$ runs from 1 to $N$.
We have seen that the field equations for the self-dual connection
derived from the complex chiral Lagrangian is compatible with those
derived from the real Lagrangian. Futhermore, we have shown that
the imaginary part of the chiral Lagrangian takes a simple form
in terms of the torsion tensor $T_{ijk}$.

For a (Majorana) Rarita-Schwinger field,
Im${\cal L}^{(+)}$ vanishes owing to the identity (\ref{Fierz}).
However, there is a possibility that Im${\cal L}^{(+)} \not= 0$
for $N = 2$ supergravity, because (\ref{Fierz}) does not hold
for this case. If this is the case, Im${\cal L}^{(+)}$ will lead
to spurious equations. (Dirac) Rarita-Schwinger fields
are also under study.



\end{document}